\documentclass{ws-procs9x6}

\usepackage{graphicx}

\begin{document}

\title{
Phase structure of NJL model
 with finite quark mass and QED correction
}

\author{Takahiro Fujihara$^{* \#}$, Tomohiro Inagaki$^{\dagger\ddagger}$,
Daiji Kimura$^\#$}
\address{
$^\#$Department of physics, Hiroshima University,
HigashiHiroshima, 739-8526, JAPAN\\
$^\ddagger$
{Information Media Center, Hiroshima University
HigashiHiroshima, 739-8521, JAPAN}\\
$^*$E-mail: fujihara@hiroshima-u.ac.jp\\
$^\dagger$E-mail: inagaki@hiroshima-u.ac.jp
}

\begin{abstract}
We study QED corrections to the chiral symmetry breaking 
 in Nambu-Jona-Lasinio (NJL) model with two flavors of quarks.
In the model the isospin symmetry is broken by differences of the current 
 quark masses and the electromagnetic charges between up and down quarks.
In the leading order of the 1/N expansion we calculate 
 the effective potential of the model with one-loop QED corrections.
Evaluating the effective potential,
 we study an influence of the isospin symmetry breaking on the 
 orientation of chiral symmetry breaking.
The current quark mass has an important contribution for
 the orientation of chiral symmetry breaking.
\end{abstract}


\section{Introduction}
The broken chiral symmetry is restored in certain
 extreme environments, e.g. high temperature, high density and so on.
The phase transition of chiral symmetry is
 understood as a non-perturbative phenomenon in QCD.
Thus the phase structure of QCD has been able to study
 in some low energy effective theories 
 or through numerical calculations in lattice QCD.
These results show that the chiral symmetry should be
 restored above a critical temperature.
It is conjectured in a lot of models that the critical temperature
 is observed blow $200\mathrm{MeV}$,
 $T_{\mathrm{cr}}< 200\mathrm{MeV}$.
Recently, an experimental study at RHIC founds an evidence
 of the symmetry restoration
 from a high temperature hadronic matter to
 deconfined partonic matter.

The effect of an electromagnetic field on chiral symmetry breaking
 is not so simple.
Low energy effective theories provide possible approaches
 for studying non-perturbative QCD phenomena in an external
 electromagnetic field.
The NJL model is one of the simplest models
 in which the chiral symmetry is broken dynamically.
The dynamical origin of symmetry breaking in the NJL model has been
 studied under an external electromagnetic field.
Then it was found that the external electromagnetic fields
 can induce a variety of phases
 \cite{Klevansky:1989vi,Suganuma:1991ha,Ishi-i:1997hv}.

Radiative QED corrections play an essential role
 for phenomena with the isospin symmetry breaking of
 quarks and hadrons, even if we consider a system with no external
 electromagnetic field.
The difference between the electromagnetic charges of up and down
 quarks breaks the SU(2) isospin symmetry.
The mass difference between up and down quarks
 also breaks the isospin symmetry.
Because of these isospin breaking,
 it is found that an interesting phase is realized inside quark
 matter through QED corrections.
If the sum of the up and down quark masses,
 $m_u+m_d$, is small enough,
 it is possible that a pion field develops a
 non-vanishing vacuum expectation value.

To study the effect of
 these isospin breaking
 we introduce terms for the current quark mass and
 QED interactions into the NJL model,
 and evaluate the effective potential
 under some assumptions.
We observe the minimum of the effective potential
 and determine the phase.
Then we evaluate the pion mass difference in our model.

\section{Gauged NJL Model}
We introduce the current quark mass
 and QED interactions into the NJL model.
This model is called the gauged NJL model.
The Lagrangian for the gauged NJL model\cite{Miransky:1988gk,Kondo:1992sq}
is
\begin{eqnarray}
 \mathcal{L}
  &=&
  \mathcal{L}_m
  + \mathcal{L}_{photon}
  + \mathcal{L}_{QED},\\
 \mathcal{L}_m
   &=&
   \bar\psi
   \left(
    i \gamma^\mu \partial_\mu -M
   \right)
   \psi
   +\frac{G}{2N}
   \left[
    (\bar{\psi}\psi)^2
    + (\bar{\psi}i \gamma_5 \tau^a\psi)^2
   \right], \\
 \mathcal{L}_{photon}
  &=&
  -\frac{1}{4} F_{\mu\nu}F^{\mu\nu}
  -\frac{1}{2\xi}(\partial_\mu A^\mu)^2, \\
 \mathcal{L}_{QED}
  &=&- \bar\psi e Q \gamma^\mu A_\mu \psi,
  \quad Q = \mathrm{diag}(2/3, -1/3).
\end{eqnarray}
Because of electric charges for up and down quarks
 are different, the isospin symmetry is explicitly broken.
We also introduce
 the isospin breaking  mass term, $M=\mathrm{diag}(m_u,m_d)$, $m_u \ne m_d$.

We evaluate the path integral for the quark fields by using the
 auxiliary field method, $\sigma\sim \bar\psi \psi$ and
 $\pi^a \sim \bar\psi i\gamma_5\tau^a \psi$.
The generating functional is given by
\begin{eqnarray}
 Z &=&
  \int \mathcal{D}\sigma\mathcal{D}\pi \mathcal{D}A
  \exp\left[
       i\int dx^4 \mathcal{L}_{photon}
       + iN\left\{
	    -\frac{1}{2G}
       	     \int d^4 x \left(\sigma^2+(\pi^a)^2\right)
	     \right. \right. \nonumber \\
 && \left. \left.
	    \vphantom{\int}
	    -i\ln\det
	    \left[
	     i\gamma^\mu(\partial_\mu + ieQA_\mu)
	     -M -\sigma -i\gamma_5\tau^a\pi^a
	    \right]
	   \right\}
      \right].
\end{eqnarray}

\begin{figure}
 \begin{center}
  \begin{minipage}[c]{0.2\textwidth}
   \includegraphics[width=\textwidth]{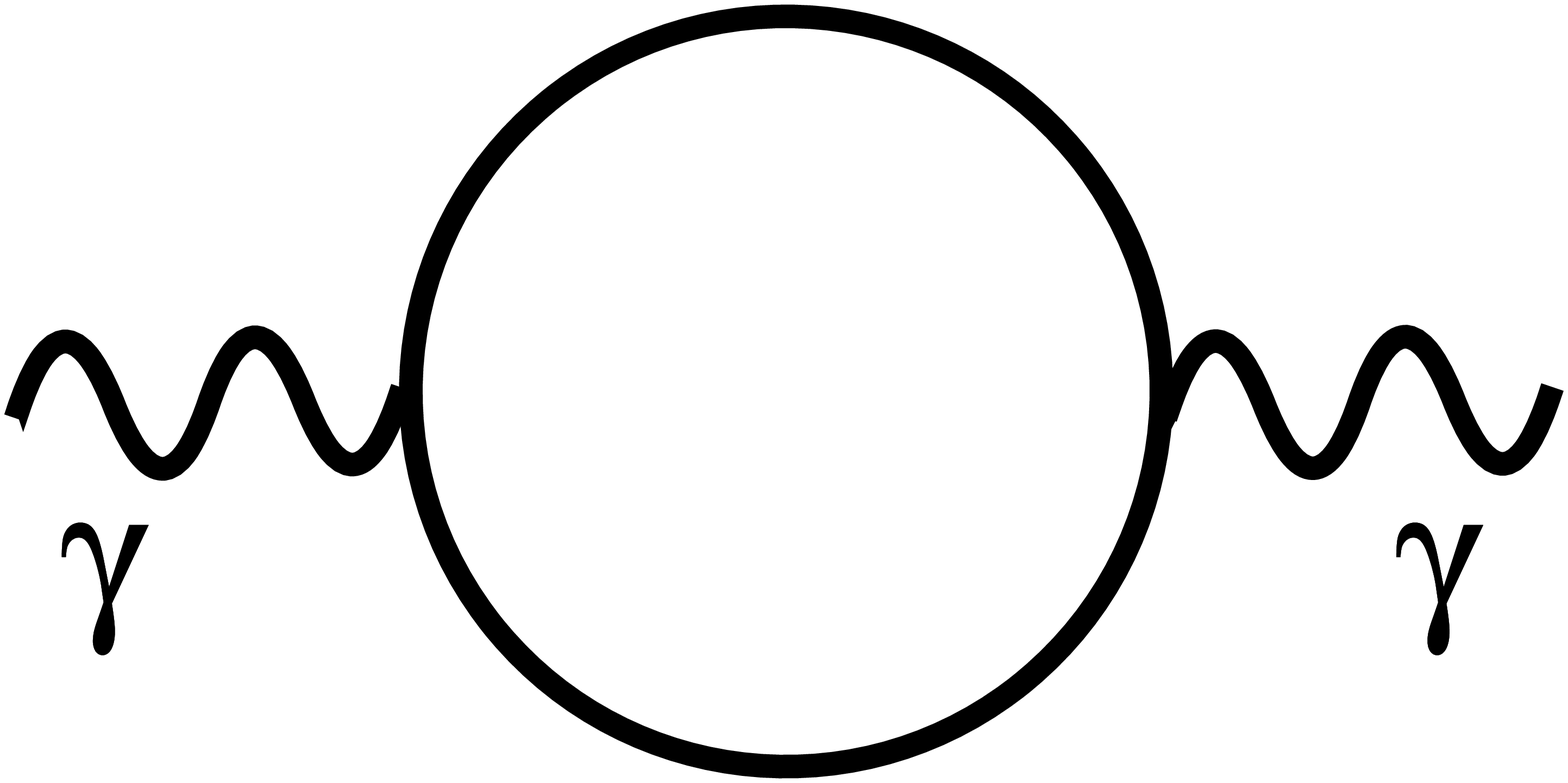}
  \end{minipage}+
  \begin{minipage}[c]{0.2\textwidth}
   \includegraphics[width=\textwidth]{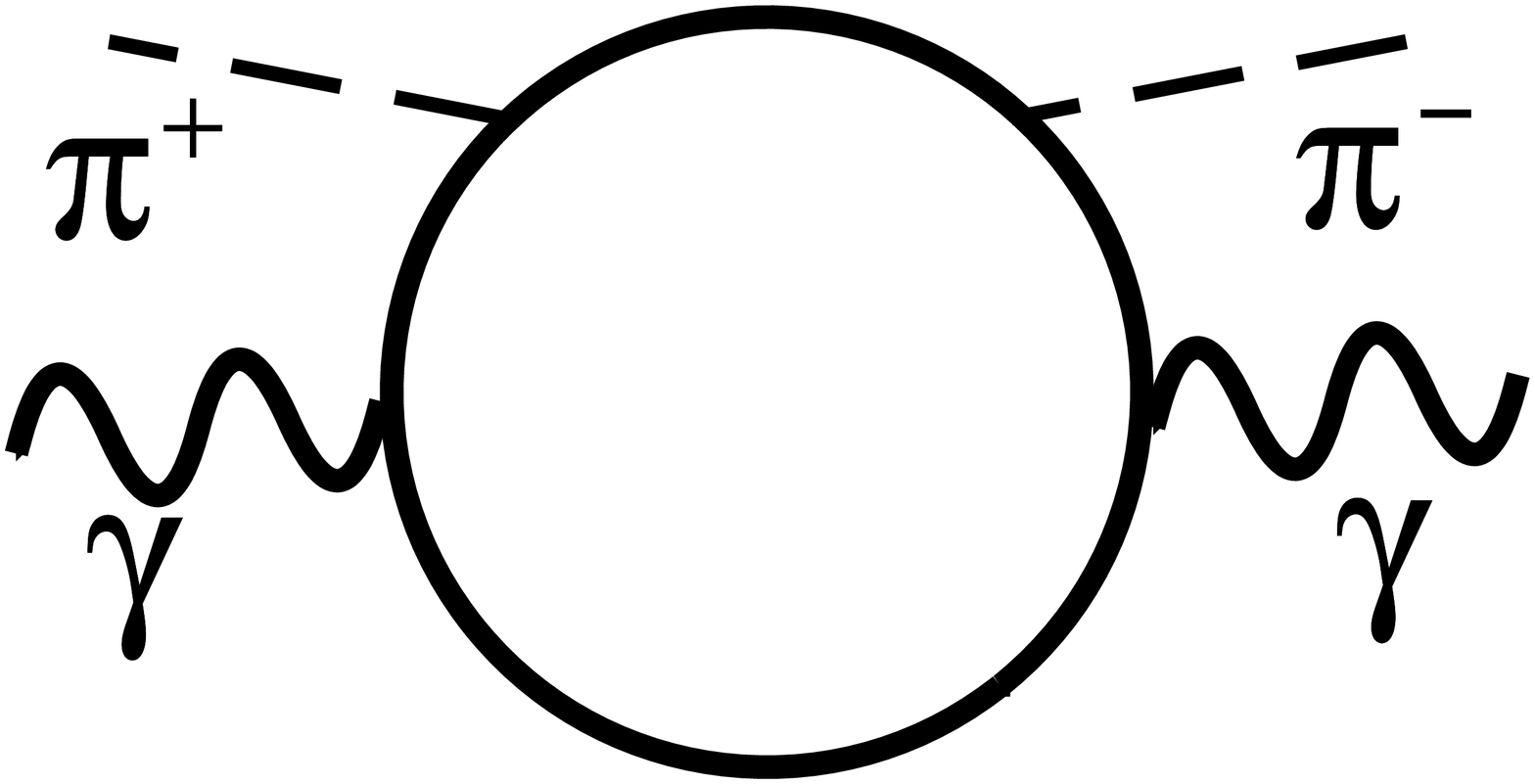}
  \end{minipage}
 \end{center}
 \caption{Diagrams of photon self-energy}
 \label{fig:photon_self}
\end{figure}

We expand the log determinant 
 in terms of the small current quark mass $M$
 and the small electric charge $e$, 
\begin{eqnarray}
&&\hspace{-5em}
i \ln \det
  \left[
   i\gamma^\mu
   \left(
   \partial_\mu
   +ieQ A_\mu
   \right)
  - M - \sigma -i \gamma_5 \tau^a \pi^a
  \right] \nonumber \\
 &=&
  i\mathrm{tr} \ln
  \left(
   i\gamma^\mu
   \partial_\mu
   \sigma -i \gamma_5 \tau^a \pi^a
  \right)
  + \mathrm{tr} \sum_{n=1}^{\infty} J_n, \\
 J_n &=&
  \frac{1}{in}
  \left(
   \frac{M+eQ\gamma^\mu A_\mu}
   {i\gamma^\mu \partial_\mu - \sigma -i \gamma_5 \tau^a \pi^a + i\epsilon}
  \right)^n.
\end{eqnarray}
We evaluate the effective potential
 up to the second order of $M$ and $e$, i.e. up to $n=2$.
As is shown in Fig.~\ref{fig:photon_self},
in this order the photon self-energy
 contributes to the effective potential.
Evaluating these diagrams in the $\overline{\mathrm{MS}}$ scheme,
we obtain the effective potential,
\begin{eqnarray}
V(\sigma,\pi^a)  &=&
V_{m}(\sigma,\pi^a) 
+ V_{gauge}(\sigma,\pi^a) ,\label{eq:veff}\\
V_m (\sigma,\pi^a) 
   &=&
   \frac{1}{4G}\sigma'{}^2
   -\frac{1}{8\pi^2}f(\sigma'{}^2;\Lambda_f^2) \nonumber \\
  && \hspace{-3em}
   -\frac{1}{4\pi^2}(m_u+m_d)\sigma g(\sigma'{}^2;\Lambda_f^2) 
   -\frac{1}{8\pi^2}(m_u^2+m_d^2)g(\sigma'{}^2;\Lambda_f^2)\nonumber \\
  &&\hspace{-3em}
   -\frac{1}{4\pi^2}
   \left[
    (m_u^2+m_d^2)\sigma^2
    + (m_u-m_d)^2 \pi^+\pi^-
   \right] \nonumber \\
  &&
   \times\left(
    \frac{\Lambda_f}{\sqrt{\Lambda_f^2+\sigma'{}^2}}
    -\ln \frac{\Lambda_f + \sqrt{\Lambda_f^2+\sigma'{}^2}}{\sqrt{\sigma'{}^2}}
   \right),\label{eq:veffm}\\
V_{gauge}(\sigma,\pi^a) 
 &=&
 \frac{1}{192\pi^2}
 \left[
  3 f\left(
      4\sigma'{}^2
      (1+4\alpha N / 27 \pi)
      - (4\alpha N /3\pi )\pi^+ \pi^-; \Lambda_p^2
     \right)
  \right. \nonumber \\
 &&
  \left.
  + f\left(
      4\sigma'{}^2
      (1+4\alpha N / 27 \pi);\Lambda_p^2
      \right)
  -4f\left(
      4\sigma'{}^2;\Lambda_p^2
      \right)
 \right],\\
 f(s^2;t^2) &\equiv&
  (2t^2+s^2)\sqrt{t^2(t^2+s^2)}-s^4\ln\frac{\sqrt{t^2}+\sqrt{t^2+s^2}}{\sqrt{s^2
}},\\
 g(s^2;t^2) &\equiv&
  \sqrt{t^2(t^2+s^2)}-s^2\ln\frac{\sqrt{t^2}+\sqrt{t^2+s^2}}{\sqrt{s^2}},
\end{eqnarray}
 where ${\sigma'}^2\equiv \sigma^2 +(\pi^a)^2$,
 $\Lambda_f$ is 3-dimensional momentum UV cutoff,
 $\Lambda_p$ is photon 3-dimensional momentum UV cutoff
 and $\alpha \equiv \frac{e^2}{4\pi^2}$.
In this paper we set $\Lambda_p = \Lambda_f$.
The ground state of the model is found by observing
 the minimum of the effective potential (\ref{eq:veff}).

Parameters in the model are determined to reproduce
 the realistic pion decay constant
 $f_\pi =91.9\mathrm{MeV}$ and 
 pion mass $m_\pi =135\mathrm{MeV}$.
For $e=0$ and $m_u=m_d$,
 the pion wave function renormalization $Z_\pi$ is given by
\begin{eqnarray}
 Z_\pi^{-1}
  &=&
  \frac{N}{2\pi^2}
  \left[
   \ln \frac{\Lambda_f + \sqrt{\Lambda_f^2+(\sigma+m_q)^2}}{\sigma+m_q}
   -\frac{\Lambda_f}{\sqrt{\Lambda_f^2+(\sigma+m_q)^2}}
  \right]. \label{eq:zpi}
\end{eqnarray}
Using Eq.(\ref{eq:zpi})
 and Gell-Mann-Oakes-Renner relation\cite{Gell-Mann:1968rz},
 we fix parameters $\Lambda=0.697\mathrm{GeV}$,
 $G=24.8\mathrm{GeV}^{-2}$ with
 $m_{u}=m_{d}=4.5\mathrm{MeV}\equiv m_q$.
Below we use these parameters
 even in cases of $m_u \ne m_d$ and $e \ne 0$.

\begin{figure}
 \begin{center}
 \begin{minipage}{0.9\textwidth}
  \includegraphics[width=\textwidth]{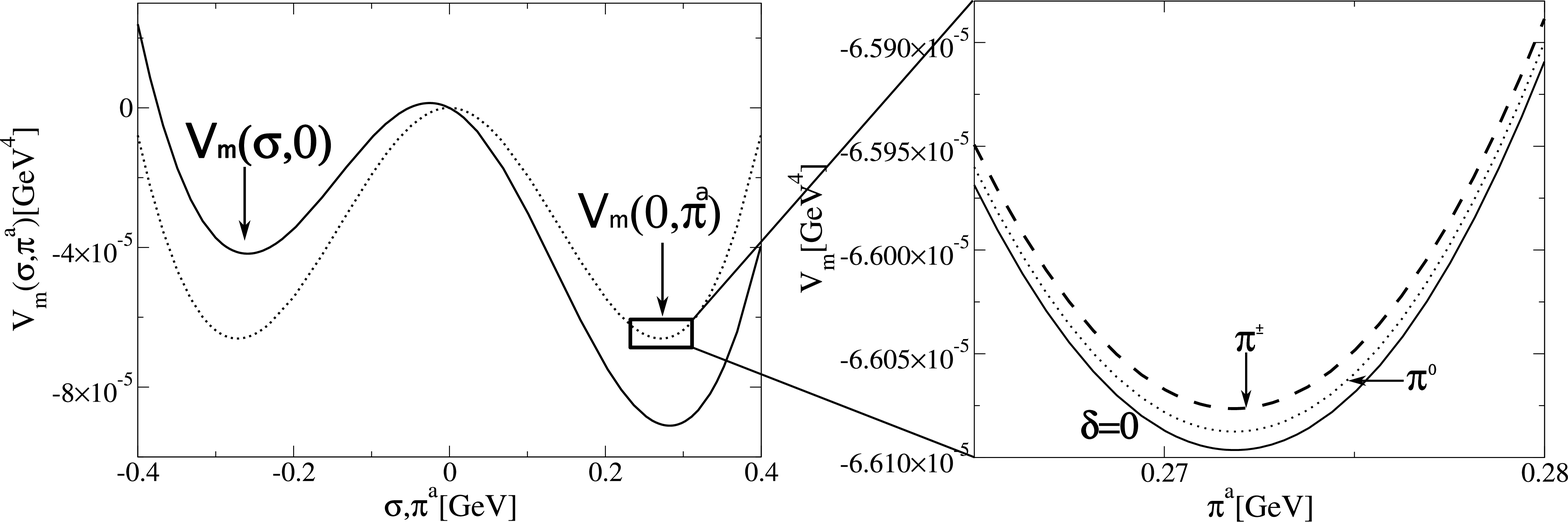}
 \caption{
  Behavior of the effective potential (\ref{eq:veffm}) without QED corrections.
  In the left figure,
  the solid line shows the effective potential
  along the sigma axis
  and the dotted line shows the one along the pion axis.
  In the right figure,
  we zoom the rectangle part in the left one.
  The solid line shows the $\delta=0$ case,
  the dotted line shows
  the effective potential along the $\pi^0$ axis
  and the dashed line shows the one along the $\pi^\pm$ axis.
  }
 \label{fig:pot_all_ex}
 \end{minipage}
 \end{center}
\end{figure}

First we evaluate the effective potential
 without QED corrections.
Evaluating the isospin breaking contribution
 from the current quark mass,
 we put
 $m_u = m_q - \delta$ for the up quark mass
 and $m_d = m_q + \delta$ for the down quark mass,
 as the sum of quark masses, $m_u+m_d$, is fixed ($m_u+m_d=9\mathrm{MeV}$).
We show the result for, $\delta= 2\mathrm{MeV}$ in Fig.~\ref{fig:pot_all_ex}.

We look for the minimum of the effective potential in
 sigma $\sigma$, neutral pion $\pi^0$
 and charged  pion $\pi^\pm$ hyper plane.
The contribution of the isospin breaking effect
 is extremely small in Fig.~\ref{fig:pot_all_ex}.
We can not distinguish the neutral pion direction and
 the charged pion direction in the left figure
 of Fig.~\ref{fig:pot_all_ex}.
The right figure of Fig.~\ref{fig:pot_all_ex} shows
 the local minimum of the effective potential
 along the pion axis.
The effect of $\delta$ splits curves
 of the effective potential for $\pi^0$ and $\pi^\pm$ a little.
We see only a small effect of the isospin breaking 
 for the difference of the dotted and the dashed lines
 in Fig.~\ref{fig:pot_all_ex}.

In this model pion mass difference is given by
\begin{eqnarray}
 \Delta m_\pi &\equiv & m_{\pi^\pm} - m_{\pi^0} \nonumber \\
  &=& \sqrt{2 N Z_{\pi}}
   \left[
    \left( \frac{\partial^2 V}
     {\partial \pi^{\pm}\partial\pi^{\pm}}\right)^{1/2}
    - \left( \frac{\partial^2 V}
       {\partial \pi^{0}\partial\pi^{0}}\right)^{1/2}
   \right] \label{eq:pi-mass-diff}.
\end{eqnarray}
For $\delta = 2\mathrm{MeV}$,
 we get the pion mass difference $\Delta m_\pi=0.06\mathrm{MeV}$.
It is experimentally observed about 5 MeV.
\begin{figure}
 \begin{center}
  \begin{minipage}[t]{0.45\textwidth}
   \includegraphics[width=\textwidth]
   {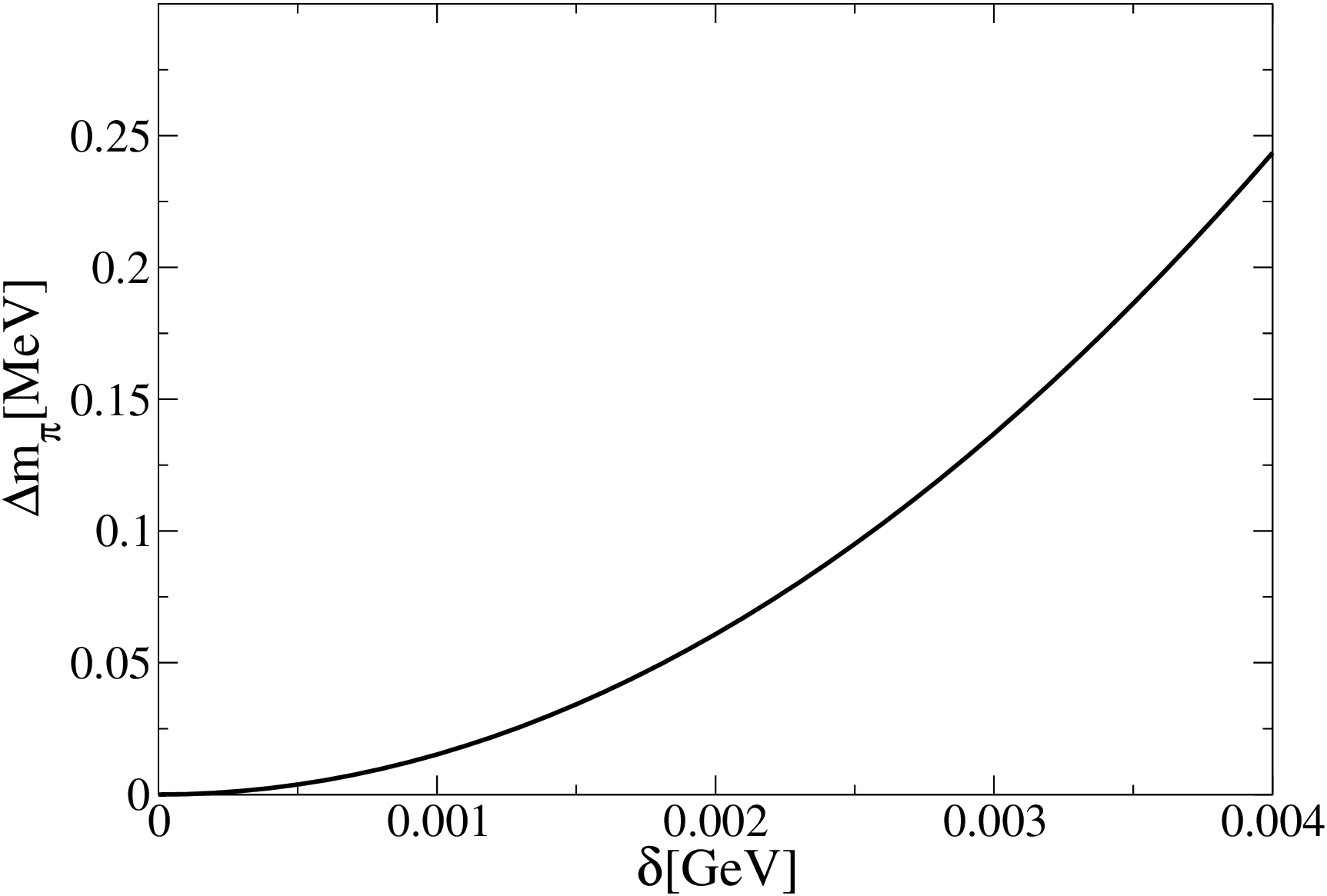}
   \caption{
   Pion mass difference as a function of $\delta$ without QED corrections.}
   \label{fig:mass-diff}
  \end{minipage}\hfill
  \begin{minipage}[t]{0.45\textwidth}
   \includegraphics[width=\textwidth]
   {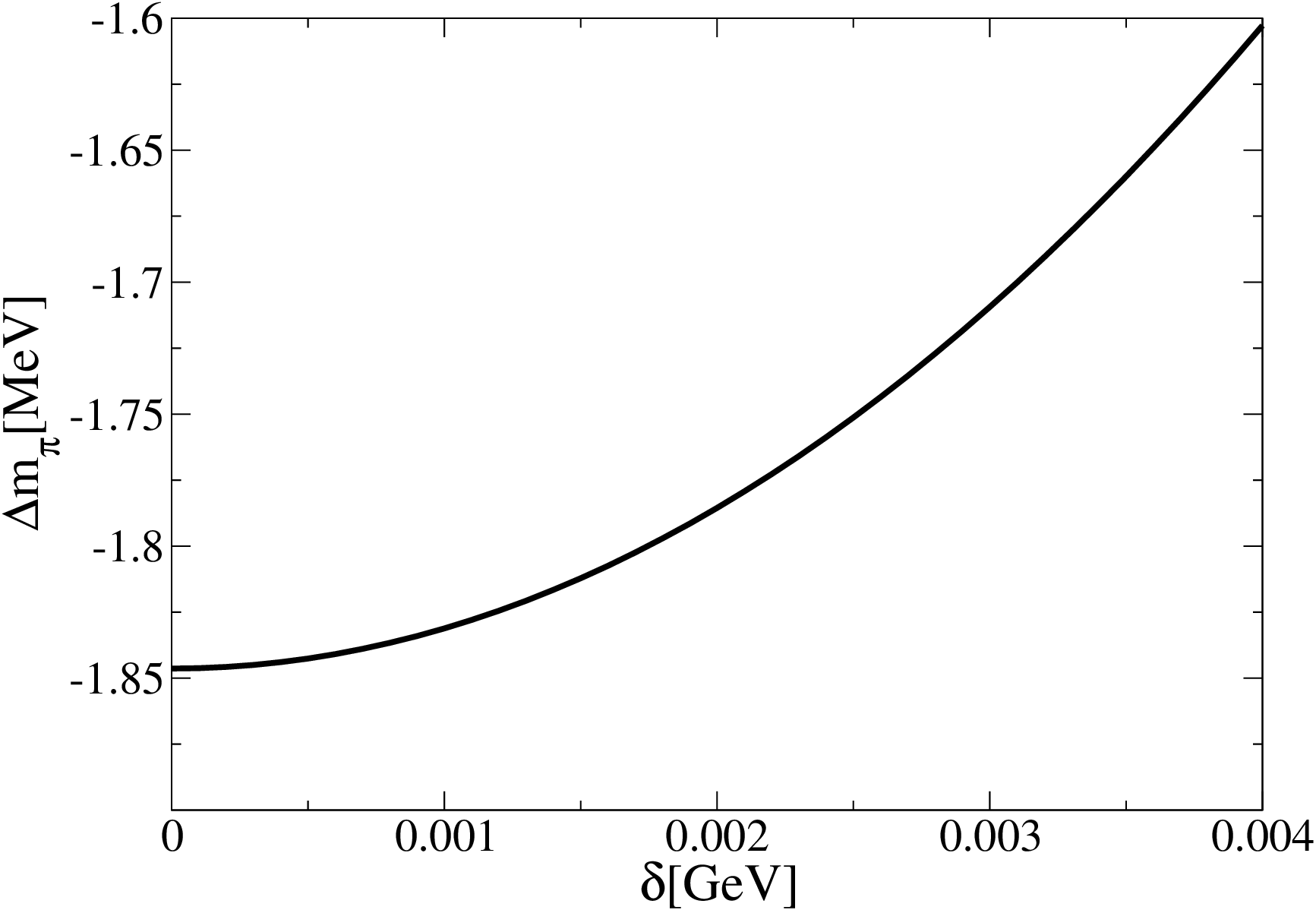}
  \caption{
   Pion mass difference as a function of $\delta$ with QED corrections.}
  \label{fig:mass-diff-e}
  \end{minipage}
 \end{center}
\end{figure}
In Fig.~\ref{fig:mass-diff}
 we plot $\Delta m_\pi$ as a function of $\delta$.
To reproduce a realistic $\Delta m_\pi$ another contribution
 have to be introduced.
 
Next we consider the QED corrections.
\begin{figure}
 \begin{center}
  \includegraphics[width=0.48\textwidth]
  {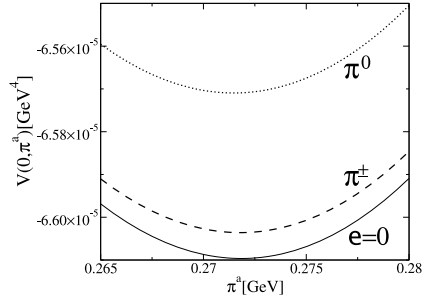}
 \end{center}
 \caption{Local minimum of the effective potential
 along the pion axis with $\delta=0$.
 The solid line shows the effective potential without QED corrections.
 The dotted line and the dashed line show the effective potential
 with QED corrections 
 along the neutral pion and the charged pion axes, respectively.}
 \label{fig:pot-pi-e1-ex}
\end{figure}
We numerically evaluate the effective potential (\ref{eq:veff})
 and illustrate the behaviors near the local minimum
 of the effective potential along the pion axis
 in Fig.~\ref{fig:pot-pi-e1-ex}.
A larger isospin breaking effect is observed in this figure.
The effective potential for
 the charged pion is smaller than that for neutral pion.
\begin{figure}
 \begin{center}
  \begin{minipage}{0.46\textwidth}
   \includegraphics[width=\textwidth]
   {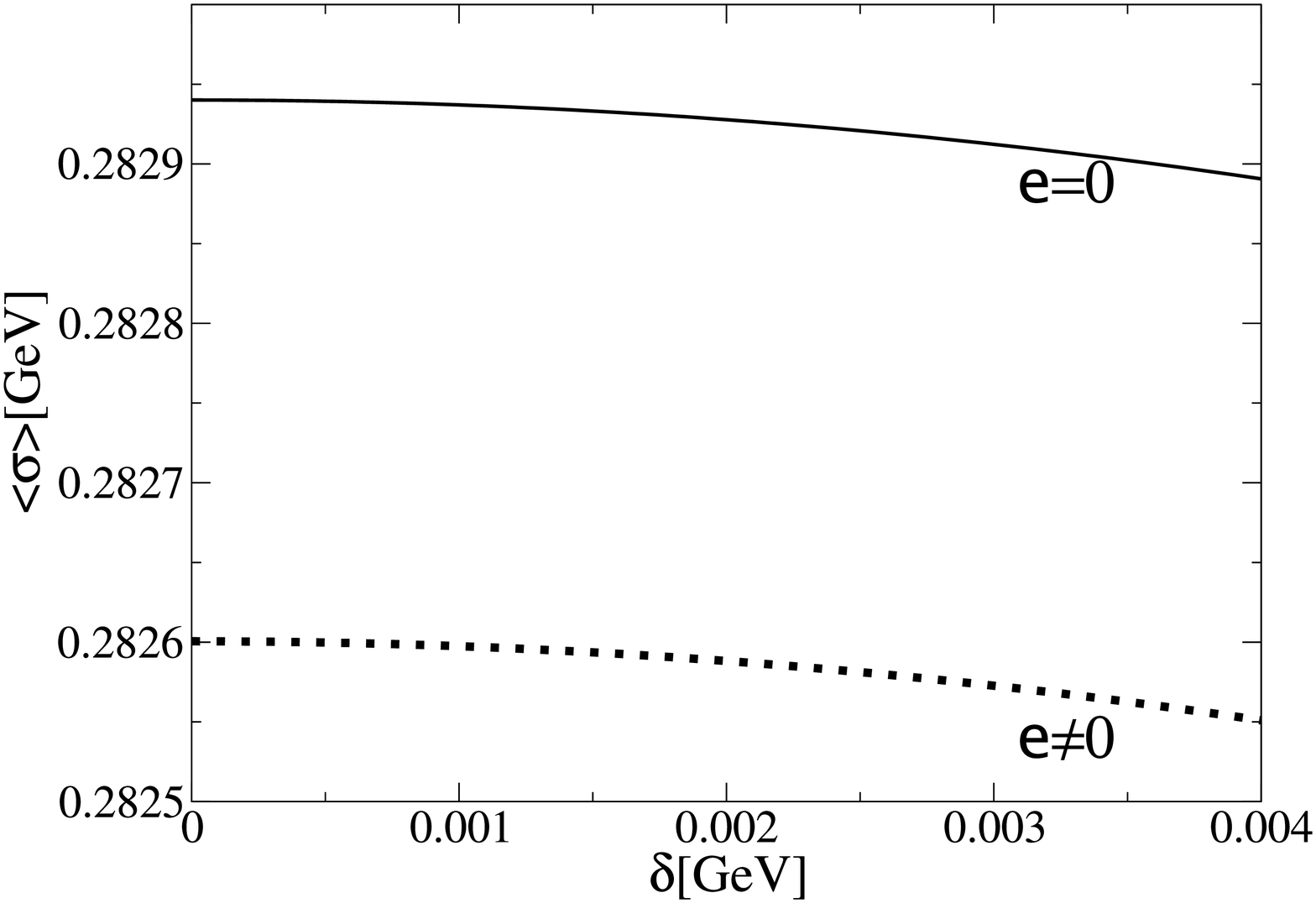}
   \caption{The global minimum of the effective potential
   as a function of $\delta$.}
   \label{fig:ev-sigma-ex}
  \end{minipage}\hfill
  \begin{minipage}{0.46\textwidth}
   \includegraphics[width=\textwidth]
   {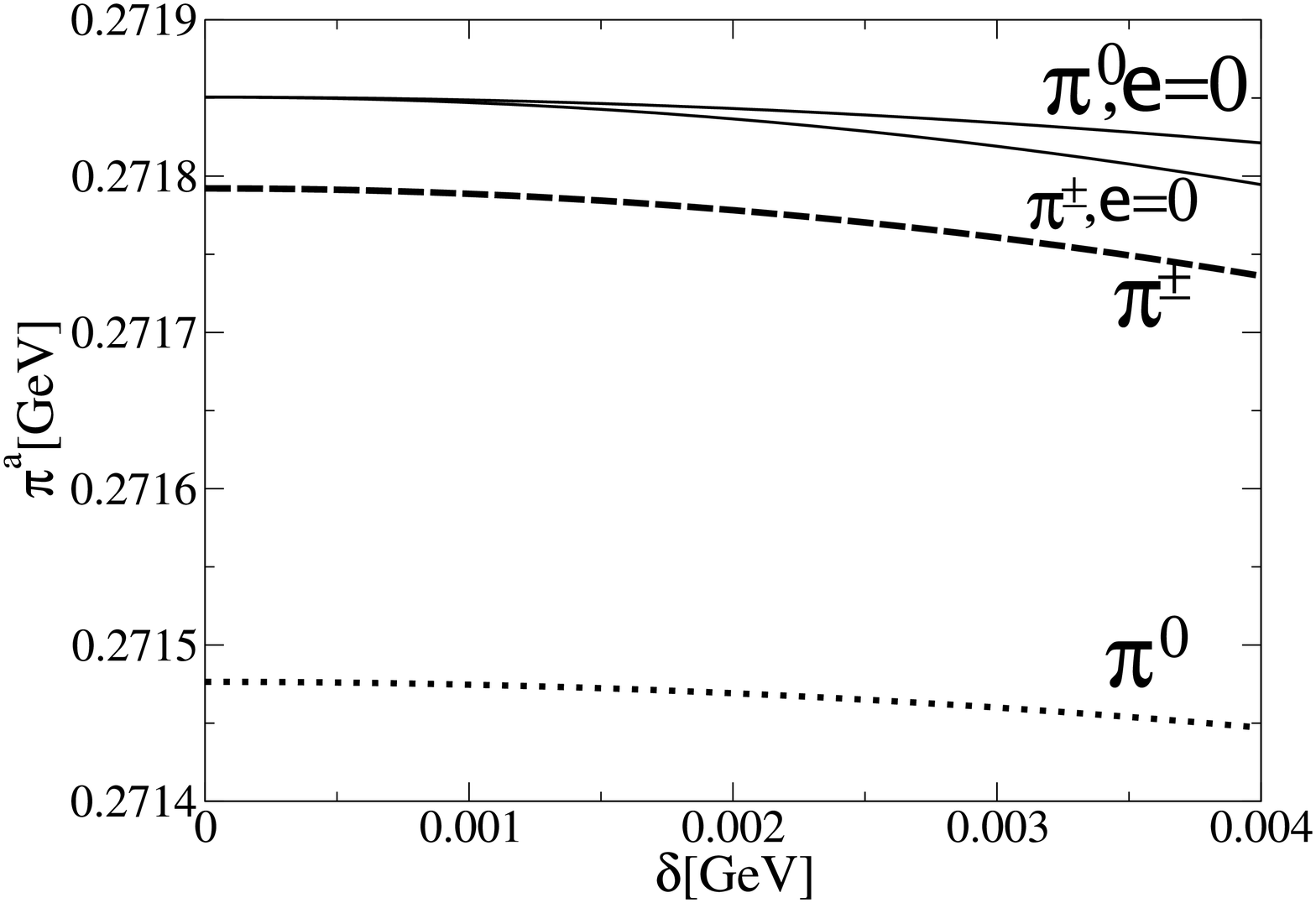}
   \caption{The local minimum of the effective potential
   along the pion axis as a function of $\delta$.}
   \label{fig:ev-pi-ex}
  \end{minipage}
 \end{center}
\end{figure}
Therefore the ground state is found at $\sigma\ne 0$ and $\pi^a=0$.
In Fig.~\ref{fig:ev-sigma-ex}
 we plot the expectation value of $\sigma$ as
 a function of $\delta$.
QED corrections decrease this expectation value,
 i.e. the chiral symmetry breaking is suppressed.

In Fig.~\ref{fig:ev-pi-ex} we show the local minimum
 of the effective potential on the pion axis.
The solid lines show the one without QED corrections.
The dotted and dashed lines mean the one with QED corrections
 along the neutral pion and the charged pion axes, respectively.
QED corrections and mass difference decrease the local minimum.
These isospin breaking split the neutral and the charged
 pion fields dependencies of the effective potential.
A contribution from the isospin breaking of
 the current quark mass slightly suppresses
 the chiral symmetry breaking.
QED corrections have much larger effect
 and also suppress the chiral symmetry breaking.

As is shown in Fig.~\ref{fig:mass-diff-e},
 the pion mass difference has a negative value.
In next section 
 we consider the phenomenological model
 to obtain the experimental pion mass difference.
\section{Phenomenological model for $\Delta m_\pi$}
As is well-known,
 a pion-photon interaction plays an important role
 in determining the pion mass difference\cite{Dmitrasinovic:1992hb}.
We introduce kinetic terms for mesons,
\begin{eqnarray}
{\cal L}& =&
 {\cal L}_{\rm m}
 + {\cal L}_{\rm photon}
 + {\cal L}_{\rm QED} + {\cal L}_{\rm s},\\
{\cal L}_{\rm s}& =& \frac1{2N} \left[
 \frac12 (\partial_\mu \sigma)^2
  + \frac12 (\partial_\mu \pi^0)^2
  + (\partial_\mu + ieA_\mu) \pi^+
    (\partial_\mu - ieA_\mu) \pi^- \right].
\end{eqnarray}
We evaluate the effective potential again
 and calculate the pion mass difference.
Then we obtain more realistic value, $\Delta m_{\pi}\sim 8\mathrm{MeV}$.

\section{Summary}
We have evaluated the effect
 of the isospin symmetry breaking due to
 current quark masses and electric charges.
The effective potential
 has been evaluated up to $O(e^2)$ and $O(m^2)$.
The isospin breaking effects split neutral and charged pion masses.
The current quark mass difference slightly suppresses the chiral
 symmetry breaking.
QED corrections also suppress the chiral symmetry breaking.
The effect of the current quark mass differences is
 extremely small for the pion mass difference.
The contribution of the QED corrections
 gives the opposite sign for the pion mass difference in our model.
To obtain a realistic pion mass difference
 we have introduced the kinetic term for meson fields.
Then we have obtained a larger pion mass difference
 $\Delta m_{\pi}\sim 8\mathrm{MeV}$.
The long distance interaction
 between photons and the charged pions
 makes the kinetic term for mesons necessary.

The temperature dependence
 of the pion mass difference
 is discussed in Ref. \refcite{Fujihara:2005wk}.

\bibliographystyle{ws-procs9x6}
\bibliography{fujihara-20061122-SCGT2006}
\end{document}